\documentclass[aps,prd,amssymb,twocolumn,superscriptaddress,floatfix,nofootinbib,10pt]{revtex4-2}
\usepackage{setspace}
\usepackage[utf8]{inputenc}
\usepackage{float}
\usepackage{amsmath,amssymb,amsfonts,bm}
\usepackage{graphicx}
\usepackage{multirow}
\usepackage[usenames,dvipsnames]{xcolor}
\allowdisplaybreaks
\usepackage[colorlinks,breaklinks,linkcolor=Blue,citecolor=Green,urlcolor=NavyBlue]{hyperref}

\usepackage{appendix}
\usepackage{orcidlink}
\usepackage{subfigure}
\usepackage{soul}
\usepackage{physics}
\usepackage{booktabs}
\usepackage{ulem}
\usepackage{tikz-feynman}

\usepackage{gentium}
\usepackage[cochineal,bigdelims,vvarbb]{newtxmath}
\usepackage[T1]{fontenc}

\newcommand{\olsi}[1]{\,\overline{\!{#1}}} % overline short italic

\usepackage[compact]{titlesec}
\titleformat{\section}
   {\centering\normalfont\bfseries\MakeUppercase}{\thesection}{1em}{}
\titlespacing*{\section}
{0pt}{3ex plus 1ex minus 1ex}{0ex plus 1ex}
\titlespacing*{\subsection}
{0pt}{2ex plus 1ex minus 1ex}{0.5ex plus 0.5ex}

\allowdisplaybreaks

\newcommand{\mev}{\textrm{\,MeV}}

\begin{document}

%\author{Jia-Xin Lin}
\author{Jia-Xin Lin\,\orcidlink{https://orcid.org/0009-0006-6382-1035}}
\email[]{linjx@seu.edu.cn}
\affiliation{School of Physics, Southeast University, Nanjing 210094, China}%
\affiliation{Instituto de F\'{\i}sica Corpuscular, Centro Mixto Universidad de Valencia-CSIC, Institutos de Investigaci\'on de Paterna, Aptdo. 22085, E-46071 Valencia, Spain}%

\author{P. Encarnación\,\orcidlink{0009-0005-0749-3885}}
\email{Pablo.Encarnacion@ific.uv.es}
\affiliation{Departamento de F\'{i}sica Teórica and IFIC, Centro Mixto Universidad de Valencia-CSIC, Institutos de Investigaci\'{o}n de Paterna, Apartado 22085, E-46071 Valencia, Spain}

%\author{Miguel Albaladejo}
\author{Miguel Albaladejo\,\orcidlink{https://orcid.org/0000-0001-7340-9235}}
\email[]{Miguel.Albaladejo@ific.uv.es}
\affiliation{Instituto de F\'{\i}sica Corpuscular, Centro Mixto Universidad de Valencia-CSIC, Institutos de Investigaci\'on de Paterna, Aptdo. 22085, E-46071 Valencia, Spain}%

%\author{Albert Feijoo}
\author{Albert Feijoo\,\orcidlink{https://orcid.org/0000-0002-8580-802X}}
\email[]{edfeijoo@ific.uv.es}
\affiliation{Instituto de F\'{\i}sica Corpuscular, Centro Mixto Universidad de Valencia-CSIC, Institutos de Investigaci\'on de Paterna, Aptdo. 22085, E-46071 Valencia, Spain}%

\title{\boldmath Signatures of Odd-Parity $s$-wave $\Xi^*$ States in Femtoscopic Correlation Functions}

\begin{abstract}
We investigate the $\Xi^*$ resonances within the molecular picture, where these states are dynamically generated as poles in the unitarized scattering amplitudes arising from the coupled-channel interactions of $K^{*-} \Lambda$, $K^{*-} \Sigma^0$, $\rho^- \Xi^0$, $\olsi{K}{}^{*0} \Sigma^-$, $\rho^0 \Xi^-$, $\omega \Xi^-$, and $\phi \Xi^-$. The interaction kernel is derived from the local hidden gauge formalism, while the unitarization procedure employs a hybrid method that combines cutoff and dimensional regularizations in the evaluation of the loop function. From a detailed spectroscopic analysis, we identify two $S = -2$ baryon states whose properties are compatible with some of the $\Xi^*$ resonances listed in the Review of Particle Physics. To explore their possible experimental signatures, we compute the femtoscopic correlation functions for all the vector--baryon pairs considered in the present study, using realistic estimates of production weights and varying source sizes $R = 1, 1.1, 1.2, 1.3, 1.5$ fm.  

\end{abstract}

\maketitle

\section{Introduction}\label{sec:Intr}

The nature and properties of hyperons, especially those of doubly and triply strange hyperons, remain a subject of considerable controversy. A total of 11 doubly strange $\Xi$ hyperons are listed by the Particle Data Group (PDG) \cite{ParticleDataGroup:2024cfk}. However, only the ground-state members of the SU(3) octet and decuplet have received a four-star rating, indicating a well-established status. The situation remains somewhat ambiguous for the $\Xi(1690)$, $\Xi(1820)$, $\Xi(1950)$, and $\Xi(2030)$ resonances: although all four are assigned a three-star status, only the $\Xi(1820)$ has an experimentally determined spin-parity, with $J^P = \frac{3}{2}^-$. In the early 1960s, the $\Xi(1950)$ state was first proposed in the decays $K^-p \to \Xi^- K^0 \pi^+$ and $K^-p \to \Xi^- K^+ \pi^-$ \cite{Badier:1965zzc},  and was later confirmed in several experiments \cite{Biagi:1981cu,Biagi:1986zj,WA89:1999nsc}. The mass and width of the $\Xi(1950)$ are reported to be $M = 1950(15) \mev$, $\Gamma = 60(20) \mev$ \cite{ParticleDataGroup:2024cfk}. In contrast, the situation becomes markedly more uncertain for the remaining $\Xi$ resonant states. For instance, the $\Xi(2120)$ state is a one-star state with a mass of approximately $2120\mev$, but its width is not experimentally known \cite{ParticleDataGroup:2024cfk}.  This state was first observed in the $\Lambda K^-$ invariant mass distribution (IMD), providing the earliest and most compelling evidence for its existence~\cite{Amsterdam-CERN-Nijmegen-Oxford:1976ezm}. Its presence was later confirmed in Ref.~\cite{French-Soviet:1979pox}. However, this state was not confirmed in the $\Lambda K^-$ IMD, despite the use of a larger event sample in the analysis~\cite{Amsterdam-CERN-Nijmegen-Oxford:1977bvi}. For a comprehensive and up-to-date overview of $\Xi$ and $\Omega$ hyperon spectroscopy, the review in Ref.~\cite{Crede:2024hur} provides a valuable resource.

The situation has progressively improved, thanks to significant advances in experimental techniques. In particular, the decays of baryons containing charm and bottom quarks have become an important source of information for deepening our understanding of the properties and internal dynamics of these hyperons. A clear example is the $\Xi^-_b \to J/\psi \Lambda K^-$ decay measured by the LHCb collaboration~\cite{LHCb:2020jpq}. This experimental breakthrough is widely recognized for providing the first evidence of a hidden-charm strange pentaquark in the $J/\psi\, \Lambda$ IMD, i. e. the $P_{\psi s}^\Lambda(4459)^0$. While the observation of this exotic state marks a turning point in hadron spectroscopy, the data collected from the $K^- \Lambda$ IMD, along with its subsequent analysis, not only improved the precision of the measured masses and widths of the $\Xi(1690)^-$ and $\Xi(1820)^-$ resonances, but also offered a unique opportunity to explore meson-baryon interactions in the strangeness $S = -2$, charge $Q = -1$ sector. The LHCb measurement was preceded by the experimental study of the $\Xi^+_c \to \pi^+\pi^+\Xi^-$ decay, conducted by Belle collaboration~\cite{Belle:2018lws}. A structure attributed to the $\Xi(1620)^0$ state was first observed in the $\Xi^-\pi^+$ IMD, enabling a precise determination of its mass and width. A distinct signal of the $\Xi(1530)^0$ resonance was also observed in the same distribution. Belle Collaboration also reported clear evidence of the $\Xi(1690)^0$ resonance in the $\olsi{K}{}^0\Lambda$ and $\Sigma^+ K^-$ IMDs observed in the $\Lambda_c^+ \to \olsi{K}{}^0\Lambda K^+(\Sigma^+ K^- K^+)$ decays~\cite{Belle:2001hyr}. Recently, a peak around the $\Xi(1820)$ region in the $K^-\Lambda$ IMD was observed in the decay $\psi(3686) \to \overline{\Xi}{}^+ K^- \Lambda$, as reported by the BESIII Collaboration~\cite{BESIII:2023mlv}. 

Ongoing theoretical efforts have progressed in parallel, aiming to deepen our understanding of the existence and nature of these $\Xi^*$ resonant states. In particular, within the effective field theory (EFT) framework, early studies employing Unitarized Chiral Perturbation Theory (UChPT) in coupled channels dynamically generated the $\Xi(1620)$~\cite{Ramos:2002xh}, followed later by the $\Xi(1690)$~\cite{Garcia-Recio:2003ejq}. In Ref.~\cite{Gamermann:2011mq}, an SU(6) extension of the chiral Lagrangian, allowing the mixing of pseudoscalar and vector mesons with octet and decuplet baryons, qualitatively described the $\Xi(1620)$, $\Xi(1690)$, and $\Xi(1950)$ as $J^P=\frac{1}{2}^-$ molecular states, as well as the $\Xi(1820)$ as a $J^P=\frac{3}{2}^-$. Further refinements using UChPT with only a contact term~\cite{Sekihara:2015qqa} and with additional leading-order (LO) diagrams~\cite{Khemchandani:2016ftn} improved the description of the $\Xi(1690)$. More recently, Ref.\,\cite{Nishibuchi:2023acl} identified the $\Xi(1620)$ just below the $\olsi{K}\Lambda$ threshold, while Ref.\,\cite{Feijoo:2023wua} included next-to-leading order (NLO) corrections for the first time in this sector. This latest study successfully reproduced both the $\Xi(1620)$ and $\Xi(1690)$ in good agreement with experimental data, and explained their decay branching ratios, finding in this way support for their molecular nature, consistent with earlier works~\cite{Sekihara:2015qqa,Khemchandani:2016ftn}.
%This latest study successfully reproduced both the $\Xi(1620)$ and $\Xi(1690)$ in good agreement with experimental data and supported their molecular nature, consistent with earlier works~\cite{Sekihara:2015qqa,Khemchandani:2016ftn}, in explaining the decay branching ratios of the $\Xi(1690)$. 

In Ref.~\cite{Sarkar:2004jh}, the interaction between pseudoscalar mesons and decuplet baryons was studied, revealing two nearby resonances, one narrow and one broad, around the $\Xi(1820)$ energy region, thus characterizing this state as a double-pole structure.\footnote{The appearance of double-pole structures due to chiral dynamics is ubiquitous. Some examples include the $\Lambda(1405)$ \cite{Oset:1997it,Oller:2000fj,Garcia-Recio:2002yxy,Jido:2003cb,Garcia-Recio:2003ejq}, the $D^{\ast}_0(2400)$ \cite{Albaladejo:2016lbb,Du:2017zvv}, and the $K_1(1270)$ \cite{Roca:2005nm,Geng:2006yb,Garcia-Recio:2010enl}. See also Refs.\,\cite{Meissner:2020khl,Xie:2023cej}.} In this context, a Weinberg–Tomozawa-like interaction was implemented within a unitarized coupled-channel framework to predict the energy levels of the pseudoscalar–decuplet system in a finite volume~\cite{Ji:2024tyo}.

The hidden gauge symmetry (HGS) formalism~\cite{Bando:1984ej,Bando:1987br,Nagahiro:2008cv}, which offers a framework to describe the dynamics of vector mesons interacting with the baryon octet, was employed to study various strange sectors in Ref.\,\cite{Oset:2010tof}. In the $S = -2$ sector, two dynamically generated poles with degenerate spin-parity $J^P = \frac{1}{2}^-$ and $\frac{3}{2}^-$ were found, whose energy positions led to their identification with the $\Xi(1950)$ and $\Xi(2120)$ resonances. A step forward was made in Ref.\,\cite{Garzon:2012np}, where additional box-like diagrams constructed with contact interaction vertices were included. These diagrams enable the coupling of external vector–baryon pairs to intermediate pseudoscalar–baryon virtual channels. Such contributions specifically affect the $J^P = \frac{1}{2}^-$ sector, breaking the original degeneracy of the states. However, their impact at the level of spectroscopy is relatively mild, resulting primarily in a slight broadening of the generated states and a small downward shift in their mass positions. The authors of Ref.~\cite{Hei:2023eqz} solved the non-relativistic Schrödinger equation using potentials derived from $t$-channel exchanges of vector and pseudoscalar mesons. One of their key findings is the possible interpretation of the $\Xi(2120)$ resonance as a vector--baryon molecular state, with spin-parity assignments $J^P = \frac{1}{2}^{\pm}$. A negative parity assignment is favored if the interaction is restricted to the $s$-wave projection.

Several of the previously discussed models have been employed to account for final-state interactions (FSIs) in theoretical analyses of the aforementioned decay processes. The works presented in Refs.~\cite{Chen:2015sxa,Feijoo:2024qqg} reproduce the experimental $K^-\Lambda$ IMD from $\Xi^-_b \to J/\psi \Lambda K^-$ decay~\cite{LHCb:2020jpq} in qualitative good agreement. Similarly, Refs.~\cite{Li:2023olv,Miyahara:2016yyh,Magas:2024mba} applied analogous theoretical frameworks to the analysis of the $\Xi^+_c \to \pi^+ \pi^+ \Xi^-$ decay~\cite{Belle:2018lws}, aiming at describing the role of FSIs in the observed spectra. Finally, in Ref.~\cite{Molina:2023uko}, the unexpectedly large width associated with the $\Xi(1820)$ peak observed in the $K^- \Lambda$ IMD from the $\psi(3686)$ decay~\cite{BESIII:2023mlv} was interpreted as a consequence of the presence of two nearby poles in the vicinity of the $\Xi(1820)$ resonance.

It is worth noting that other theoretical approaches have addressed the $S=-2$ baryon spectroscopy from a conventional perspective, interpreting the $\Xi^*$ states as three-quark configurations. Among them, it should be stressed the studies based on conventional quark models \cite{Chao:1980em,Capstick:1986ter,Glozman:1995fu,Glozman:1997ag,Bijker:2000gq,Valcarce:2005rr,Arifi:2022ntc,Pervin:2007wa,Xiao:2013xi,Menapara:2021dzi}, large-$N_c$ analysis \cite{Schat:2001xr,Goity:2003ab,Matagne:2004pm,Matagne:2006zf,Semay:2007cv}, others employing QCD sum rules \cite{Jido:1996zw,Lee:2002jb,Aliev:2018hre} or lattice QCD \cite{Engel:2013ig} and calculations with a Skyrme model \cite{Oh:2007cr}. In Ref.~\cite{PavonValderrama:2011gp}, the authors presented theoretical arguments, based on the Gell-Mann–Okubo mass relation, in favor of interpreting the $\Xi(1950)$ resonance as a superposition of several nearby states, including members of the $J^P = \frac{1}{2}^-$ decuplet and the $J^P = \frac{5}{2}^{\pm}$ octet. This hypothesis was also supported by the authors of Ref.~\cite{Xiao:2013xi}, albeit from the perspective of a chiral quark model. Additionally, a recent study~\cite{Yan:2024usf} employing the quark delocalization color screening model found a $\Lambda \olsi{K}{}^*$ bound state with $J^P = \frac{1}{2}^-$, which is compatible with the $\Xi(1950)$ resonance.

Over the past decades, hadron femtoscopy has emerged as a leading technique for providing novel constraints on hadron interactions, particularly in sectors where traditional scattering experiments are not technically feasible. In other words, correlation functions (CFs) have demonstrated their potential to extract information on hadron interactions, scattering parameters, and reveal the presence of hadron resonances and bound states. Experimental measurements involving hadron--hadron interactions with strange quark content are already available in Refs.~\cite{STAR:2014dcy,ALICE:2017jto,ALICE:2019gcn,STAR:2018uho,ALICE:2018ysd,ALICE:2019hdt,ALICE:2020wvi,ALICE:2021cpv,ALICE:2021szj,ALICE:2021njx,ALICE:2022yyh,ALICE:2023eyl}. Building on recent experimental results, theoretical studies are also progressing, with growing efforts to model the underlying hadron interactions and to interpret femtoscopic correlation data~\cite{Morita:2014kza,Morita:2016auo,Ohnishi:2016elb,Hatsuda:2017uxk,Haidenbauer:2018jvl,Mihaylov:2018rva,Kamiya:2021hdb,Liu:2023uly,Li:2023pjx,Li:2024tvo,Molina:2023jov,Ikeno:2023ojl,Encarnacion:2024jge,Feijoo:2023wua,Feijoo:2024bvn,Albaladejo:2025lhn,Albaladejo:2025kuv}. In the context of the present study, it is worth highlighting the recent measurement of the $K^- \Lambda$ CF by the ALICE collaboration at the LHC~\cite{ALICE:2023wjz}. The measurement was theoretically analyzed in Ref.~\cite{Sarti:2023wlg}, marking the inception of a novel method to extract hadron–hadron scattering amplitudes. Notably, the low-energy constants (LECs) of the NLO chiral Lagrangian in the $S=-2$ sector were determined for the first time using femtoscopic data. In addition to its theoretical significance as the first measurement involving vector mesons, the recent measurement of the $\phi p$ CF~\cite{ALICE:2021cpv} provides further motivation by presenting the first experimental evidence for an attractive $\phi p$ interaction.
 Using the Lednicky-Lyuboshitz approximation, the extracted $\phi p$ scattering length indicated a predominantly elastic interaction.
 In contrast, the authors of Ref.~\cite{Feijoo:2024bvn} obtained a scattering length more consistent with previous results. Their theoretical approach is based on a unitary extension of the HGS formalism fitted to $\phi p$ CF data, emphasizing the crucial role of coupled-channel dynamics in femtoscopy. In addition, a near-threshold, degenerate $J^P = (1/2^-, 3/2^-)$ molecular state was also identified, much closer to the $\phi p$ threshold than in earlier works \cite{Oset:2010tof,Garzon:2012np}. A subsequent study~\cite{Abreu:2024qqo} conducted a similar analysis, incorporating additional leading-order diagrams and mixing with pseudoscalar mesons. Reasonable agreement with ALICE data was achieved by combining the two spin channels into a spin-averaged observable.

 With this renewed background and in light of the ongoing measurement of the $\rho^0p$ CF by ALICE, it seems natural to extend theoretical femtoscopy studies involving vector–baryon pairs to sectors with non-zero strangeness. A recent work~\cite{Encarnacion:2025luc} has taken a step forward in this direction by studying the $S=-1$ baryon spectroscopy from vector–baryon interactions within the femtoscopy framework. In the present study, we focus on the $\Xi^*$ resonances from a molecular perspective, which can be generated from the vector-baryon (VB) interaction derived within the HGS formalism following a hybrid unitary scheme in coupled channels, which combines the cutoff and dimensional regularization methods in the loop function. With the aim of enabling future comparisons with observables to further elucidate the existence and properties of the found poles, we calculate the CFs of the coupled channels in our scheme to identify possible traces of the resonance structure. To obtain more realistic estimates of such CFs, we also compute the production weights for high-multiplicity events employing the production yields obtained using the Thermal-FIST package and following the VLC Method~\cite{Encarnacion:2024jge}. Beyond shedding light on the odd-parity $\Xi^*$ baryon spectrum, the present study may contribute to establishing a strategy for improving EFTs by constraining LECs using femtoscopy data.

This paper is organized as follows.  Theoretical formalism is described in detail in Sec.~\ref{sec:forma}. The numerical results are presented in Sec.~\ref{Sec:res}. Finally, a brief conclusion is given in Sec.~\ref{Sec:con}.

\section{formalism}\label{sec:forma}
\subsection{\boldmath $VB$ interactions in the $S=-2$, $Q=-1$ sector}

This study is restricted to the $S=-2$, $Q=-1$ sector, as the primary goal is to investigate $\Xi^*$ spectroscopy arising from the interaction between vector mesons and ground-state baryons. Including the $S=-2$, $Q=0$ sector would provide information on the neutral counterparts of the states found in the $Q=-1$ sector. However, studying this additional neutral sector within femtoscopy framework would require the incorporation of Coulomb interactions, significantly increasing the complexity of the CF calculations without offering fundamentally new insights beyond those already accessible from the $Q=-1$ sector. Therefore, by focusing exclusively on the $S=-2$, $Q=-1$ sector, the number of relevant channels amounts to seven, namely (with the corresponding threshold in MeV): $K^{*-} \Lambda(2007)$, $K^{*-} \Sigma^0(2084)$, $\rho^-\Xi^0(2090)$, $\olsi{K}{}^{*0} \Sigma^-(2093)$, $\rho^0 \Xi^-(2097)$, $\omega \Xi^-(2104)$ and $\phi \Xi^-(2341)$.

To study the interactions of this system, we consider the exchange of vector mesons in a $t$--channel diagram, whose $VVV$ and $VBB$ vertices can be derived from the following Lagrangians
\begin{subequations}\label{eq:lagrangians}
\begin{align}
    \label{eq:VVV}
    \mathcal{L}_{VVV} &= ig \langle (V^\nu \partial_\mu V_\nu - \partial_\mu V^\nu V_\nu) V^\mu \rangle, \\[2mm]
    \label{eq:VBB}
    \mathcal{L}_{VBB} &= g\left( \langle \bar{B} \gamma_\mu [V^\mu, B] \rangle + \langle \bar{B}\gamma_\mu B \rangle \langle V^\mu \rangle \right),
\end{align}
\end{subequations}
where the symbol $\langle \cdots \rangle$ represents the trace in flavour space, $g = M_V/(2f_\pi) ~ (M_V = 800 \mev, f_\pi = 93 \mev)$, and $V_\mu$ and $B$ are matrices for the vector-meson and baryon fields (see more details in Refs.~\cite{Oset:2010tof,Feijoo:2024bvn} and references therein).

For the vector octet-baryon octet interaction studied in this work, states with quantum numbers\footnote{Note that the $J^P={\frac{1}{2}}^-$ and ${\frac{3}{2}}^-$ kernels obtained from Eqs.\,\eqref{eq:lagrangians} are degenerate. Therefore, we only need to calculate the ${\frac{1}{2}}^{-}$ ones.} $J^P={\frac{1}{2}}^{-}$ can be formed with orbital angular momentum $L$ and spin $S$ combinations $(L,S)=(0,\frac{1}{2})$ and $(2,\frac{3}{2})$. We neglect the contribution of $D$-waves as a low-energy approximation. In the same way, the projection of interaction kernel arising from the $t$-channel diagram is obtained in the non-relativistic approximation. Furthermore, we consider $\left\lvert t \right\rvert \ll M_V^2$ in the $t$--channel exchange diagrams, so that the amplitudes effectively reduce to contact interactions. The resulting Weinberg--Tomozawa-type kernel can be expressed in a relativistic form as 
\begin{eqnarray}
V_{ij} &=& -\frac{1}{4f^2} C_{ij} \sqrt{\frac{M_i + B_i}{2M_i}} \sqrt{\frac{M_j + B_j}{2M_j}} (2\sqrt{s} - M_i - M_j) \,,
\label{eq:Vij}
\end{eqnarray}
where the coefficients $C_{ij}$ are shown in Table~\ref{Tab:cij}, $M_i \,(M_j)$ and $B_i \, (B_j)$ are the initial (final) masses and energies of the baryons, respectively, and $\sqrt{s}$ is the total energy of the $VB$ system in the center-of-mass (CM).
\begin{table}[t]
\centering
\caption{Coefficients $C_{ij}$ of Eq.~\eqref{eq:Vij}.}
\setlength{\tabcolsep}{2pt}
\begin{tabular}{cccccccc}
\hline\hline
& $K^{*-} \Lambda$ & $K^{*-} \Sigma^0$ & $\rho^-\Xi^0$ & $\olsi{K}{}^{*0} \Sigma^-$ & $\rho^0 \Xi^-$ & $\omega \Xi^-$ & $\phi \Xi^-$ \\
\hline
$K^{*-} \Lambda$ & $0$ & $0$ & $-\sqrt{\frac{3}{2}}$ & $0$ & $-\frac{\sqrt{3}}{2}$ & $-\frac{\sqrt{3}}{2}$ & $\sqrt{\frac{3}{2}}$ \\
$K^{*-} \Sigma^0$ &  & $0$ & $\frac{1}{\sqrt{2}}$ & $\sqrt{2}$ & $-\frac{1}{2}$ & $-\frac{1}{2}$ & $\frac{1}{\sqrt{2}}$ \\
$\rho^-\Xi^0$ &  &  & $1$ & $0$ & $\sqrt{2}$ & $0$ & $0$ \\
$\olsi{K}{}^{*0} \Sigma^-$ &  &  &  & $1$ & $\frac{1}{\sqrt{2}}$ & $-\frac{1}{\sqrt{2}}$ & $1$ \\
$\rho^0 \Xi^-$ &  &  &  &  & $0$ & $0$ & $0$ \\
$\omega \Xi^-$ &  &  &  &  &  & $0$ & $0$ \\
$\phi \Xi^-$ &  &  &  &  &  &  & $0$ \\
\hline\hline
\end{tabular}
\label{Tab:cij}
\end{table}

Nonperturbative schemes prevent the divergence of purely perturbative EFTs and, by construction, ensure the unitarity and analyticity of the scattering amplitude. In this work, following the pioneering approach of Refs.~\cite{Oller:1997ti,Oset:1997it}, unitarity is implemented by solving the Bethe--Salpeter equation (BSE)  in coupled channels. Once the interaction kernel $V_{ij}$ is obtained, the BSE is solved by factorizing both the kernel and the scattering amplitude outside the integral. This procedure reduces the problem to a set of algebraic equations, leading to:
\begin{equation}
\label{eq:T}
T = \left[ 1 - VG \right]^{-1} V \,,
\end{equation}
where $T$ is the scattering amplitude, and $G$ is a diagonal matrix with elements given by the loop function as follows,
\begin{equation}
    G_j(s) = i \int \frac{\mathrm{d}^4 q}{(2\pi)^4} \frac{2M_j}{(P - q)^2 - M_j^2 + i\epsilon} \frac{1}{q^2 - m_j^2 + i\epsilon},
\end{equation}
with $M_j$ and $m_j$ the masses of the vector meson and baryon in the loop function and $P$ the four momentum of the $VB$ system. Since this loop function is divergent, it can be regularized either using the cutoff scheme or the dimensional regularization scheme.

The cutoff scheme [$G_\Lambda(s)$] is determined by a parameter $\Lambda$, with dimensions of energy, that can be related to the range of the interaction in momentum space. However, this method has the disadvantage that it introduces an unwanted and unphysical branch point at the values where the on shell momentum of the vector equals the value of $\Lambda$. This can be particularly problematic for the cases in which one has coupled channels spanning a large energy gap between thresholds.
The dimensional regularization scheme [$G_{\rm DR}(s)$] contrasts with the previous approach. While it preserves good analytical behavior, the dimensional regularization scheme depends on the subtraction constant $a(\mu)$, whose physical interpretation is less transparent than that of a momentum cutoff. Therefore, we adopt a hybrid approach that combines both the cutoff and dimensional regularization schemes, defining the loop function $G_j(s)$ as:
\begin{equation}
\label{eq:G}
G_j(s)=G_\Lambda^{(j)}(s_{\text{th}}) + [G_{\rm DR}^{(j)}(s) - G_{\rm DR}^{(j)}(s_{\text{th}})]\,,
\end{equation}
where the explicit forms for the functions $G_\Lambda^{(j)}(s)$ and $G_\text{DR}^{(j)}(s)$ are taken from Refs.\,\cite{Oller:1998hw} and \cite{Oller:1998zr}, respectively. Note, however, the extra factor $2M_j$ included in our definition, which arises from the normalization of the spinors associated with the corresponding $j$-th baryon. The final expression for $G_j(s)$ can be written as
\begin{widetext}
\begin{eqnarray}
    G_j(s) & = &\frac{1}{4\pi^2} \frac{M_j}{m_j+M_j} \left
(m_j\ln\frac{m_j}{\Lambda + \sqrt{\Lambda^2+m_j^2}}+  M_j\ln\frac{M_j}{\Lambda + \sqrt{\Lambda^2+M_j^2}} \right)  +  \frac{2M_j}{16\pi^2} \frac{M_j-m_j}{M_j+m_j}\left( \frac{(M_j+m_j)^2}{s}-1\right) \ln \frac{M_j}{m_j}\nonumber\\
  &&+ \frac{M_j\sigma_j}{16\pi^2 s} \left\{ \ln\left( s - M_j^2 + m_j^2 + \sigma_j) \right) - \ln\left( -s + M_j^2 - m_j^2 + \sigma_j \right) + \ln\left( s + M_j^2 - m_j^2 + \sigma_j \right) - \ln\left( -s - M_j^2 + m_j^2 + \sigma_j \right) \right\}\,,
\label{eq:loop_hybrid}
\end{eqnarray}
\end{widetext}
where $\sigma_j(s)=2\sqrt{s}p_j(s)$ and $p_j(s)$ is the CM momentum of the channel pair at an energy $\sqrt{s}$, $p_j(s)=\lambda^{1/2}(s, m_j^2, M_j^2)/(2\sqrt{s})$, with $\lambda$ the K\"all\'en function $\lambda(a, b, c) = a^2 + b^2 + c^2 - 2(ab + bc + ca)$. In this hybrid way, the loop function $G_j(s)$ is determined by a cutoff value $\Lambda$ rather than a subtraction constant $a(\mu)$, since the combination $a(\mu) + \log \frac{M_j m_j}{\mu^2}$ cancels in the difference $G_\text{DR}^{(j)}(s) - G_\text{DR}^{(j)}(s_\text{th})$. From a theoretical point of view, the function $G_{j}(s)$ in Eq.~\eqref{eq:G} follows from a once-subtracted dispersion relation, in which one writes the subtraction constant in terms of a cutoff $\Lambda$ rather than the usual $a(\mu) + \log \frac{M_j m_j}{\mu^2}$ combination. We search for poles in the second Riemann sheet:
\begin{equation}
    \label{eq:G2}
    G^{\text{II}}_j(s) = G_j(s) + i \frac{2M_j}{4\pi\sqrt{s}} p_j(s)\,,
\end{equation}
for channels where Re$\sqrt{s} > M_j + m_j$.

The results obtained from the formula discussed above are based on the approximation that the vector mesons are stable, neglecting their finite widths. Following the approach of Refs.~\cite{Oset:2010tof, Feijoo:2024bvn}, to achieve more realistic results, we incorporate the finite widths of the $\olsi{K}{}^*$ and $\rho$ mesons by folding the loop functions with their corresponding spectral functions:
\begin{eqnarray}
    \label{eq:G_tilde}
    \widehat{G}_j(s) &=& \frac{1}{N_j} \int_{\left(m_j-2 \Gamma_j\right)^2}^{\left(m_j+2 \Gamma_j\right)^2} d m^2  \operatorname{Im}\left[\frac{1}{m^2-m_j^2 + i m_j \Gamma_j(m)}\right] \nonumber \\
    && \times \, G_j\left(s, m^2, M_j^2\right)\,,
\end{eqnarray}
with the normalization factor $N_j$:
\begin{equation}
    \label{eq:N}
    N_j=\int_{\left(m_j-2 \Gamma_j\right)^2}^{\left(m_j+2 \Gamma_j\right)^2} d m^2  \operatorname{Im}\left[\frac{1}{m^2-m_j^2+i m_j \Gamma_j(m)}\right],
\end{equation}
and the energy-dependent width $\Gamma_j(m)$ for the $\olsi{K}{}^*$ and $\rho$ mesons given by:
\begin{equation}
    \label{eq:Gamma}
    \Gamma_j(m)=\Gamma_j \frac{m_j^2}{m^2} \left( \frac{q_j(m^2)}{q_j(m_j^2)} \right)^{3} \theta\left(m-\left(\overline{m}_{1(j)}+\overline{m}_{2(j)}\right)\right),
\end{equation}
where the decay widths $\Gamma_j$ of $\olsi{K}{}^*$ and $\rho$ mesons are $48.3\mev$ and $149.77\mev$, respectively, and the decay products are $\overline{m}_{1(j)} = m_{\olsi{K}}$, $\overline{m}_{2(j)} = m_\pi$ for the $\olsi{K}{}^*$ meson, and $\overline{m}_{1(j)} = \overline{m}_{2(j)} = m_\pi$ for the $\rho$ meson. Above, $q_j(m^2)$ is the momentum of the decay products in their center-of-mass frame, $q_j(m^2) =\lambda^{1/2}(m^2,\overline{m}_{1(j)}^2,\overline{m}_{2(j)}^2)/(2m)$. In practical terms, the $G_j(s)$ functions [Eq.\,\eqref{eq:loop_hybrid}] are replaced in Eq.\,\eqref{eq:T} with their convoluted versions, Eq.\,\eqref{eq:G_tilde}.

It is important to note that a difficulty arises when searching for a pole in the complex plane, as the widths of unstable mesons, such as the $\rho$ and $\olsi{K}{}^*$, smear the fixed thresholds into mass distributions. This, in turn, leads to a blurred interpretation of the threshold itself and complicates the strict definition of Riemann sheets. We can estimate the effect of the widths of the unstable mesons $\olsi{K}{}^*$ and $\rho$ in the system using the method suggested in Ref.~\cite{Oset:2010tof}. The pole positions $M_R$ and the widths $\Gamma$ are obtained from the amplitudes on the real axis, where $M_R$ and $\Gamma$ correspond to the position of the maximum of $|T|^2$ and the width at half-maximum of this magnitude, respectively.

The couplings $g_i$ of the resonances associated with the poles to the $i$-th $VB$ channel can be obtained from the residue of the $T_{ij}$ amplitude at the pole position.
Based on the mass and width obtained on the real axis, we follow the procedure of Ref.~\cite{Oset:2010tof} and obtain the couplings from the amplitudes on the real axis as follows, assuming that these amplitudes behave as $T_{ij}(\sqrt{s})=g_i\, g_j/(\sqrt{s} - M_R + i\Gamma/2)$. Then, one finds $\left| g_i \right|^2 = \Gamma \, \left| T_{ii}(\sqrt{s} = M_R) \right| / 2$, which allows for the determination of the coupling $g_i$ of the channel that couples most strongly to the resonance, up to a global phase. Similarly, the couplings associated with the other channels can be obtained from $g_j = g_i\,T_{ij}(\sqrt{s} = M_R)/T_{ii}(\sqrt{s} = M_R)$.

\subsection{Correlation function}
\begin{table*}[!t]
\centering
\caption{Values of the production weight for $K^{*-} \Lambda, K^{*-} \Sigma^0, \rho^-\Xi^0, \olsi{K}{}^{*0} \Sigma^-, \rho^0 \Xi^-, \omega \Xi^-, \phi \Xi^-$ CFs. These values are obtained following the VLC method (see a detailed explanation of the methodology in Appendix~A of Ref.~\cite{Encarnacion:2024jge}).}
\setlength{\tabcolsep}{4pt}
{%\small%
\begin{tabular}{llccccccc}
\hline\hline
\multicolumn{2}{c}{channel$-j$} & $K^{*-} \Lambda$ & $K^{*-} \Sigma^0$ & $\rho^-\Xi^0$ & $\olsi{K}{}^{*0} \Sigma^-$ & $\rho^0 \Xi^-$ & $\omega \Xi^-$ & $\phi \Xi^-$ \\
\hline
$[K^{*-} \Lambda ~ \mathrm{CF}] $ & $ \omega_j$ & $1$ & $0.043(3)$ & $0.062(5)$ & $0.036(3)$ & $0.032(3)$ & $0$ & $0$ 
\\
\hline
$[K^{*-} \Sigma^0 ~ \mathrm{CF}] $ & $ \omega_j$ & $1.78(16)$ & $1$ & $0.97(9)$ & $0.92(8)$ & $0.84(8)$ & $0.68(6)$ & $0$ 
\\
\hline
$[\rho^-\Xi^0 ~ \mathrm{CF}] $ & $ \omega_j$ & $1.87(16)$ & $0.98(8)$ & $1$ & $0.95(8)$ & $0.87(8)$ & $0.70(6)$ & $0$ 
\\
\hline
$[\olsi{K}{}^{\ast 0} \Sigma^- ~ \mathrm{CF}] $ & $ \omega_j$ & $1.79(15)$ & $1.03(9)$ & $0.98(9)$ & $1$ & $0.91(8)$ & $0.71(7)$ & $0$ 
\\
\hline
$[\rho^0 \Xi^- ~ \mathrm{CF}] $ & $ \omega_j$ & $1.90(18)$ & $1.14(11)$ & $1.08(11)$ & $1.06(9)$ & $1$ & $0.79(7)$ & $0$ 
\\
\hline
$[\omega \Xi^- ~ \mathrm{CF}] $ & $ \omega_j$ & $2.02(19)$ & $1.26(11)$ & $1.17(11)$ & $1.22(10)$ & $1.13(11)$ & $1$ & $0$ 
\\
\hline
$[\phi \Xi^- ~ \mathrm{CF}]$ & $\omega_j$ & $6.97(70)$ & $5.34(52)$ & $4.98(48)$ & $5.17(50)$ & $4.85(46)$ & $4.97(47)$ & $1$ \\
\hline\hline
\end{tabular}%
}%
\label{Tab:weight}
\end{table*}

Experimentally, the two-particle CF is obtained as the ratio of the relative momentum distribution of pairs of particles produced in the same event and a reference distribution of pairs originated in different collisions.
Within certain approximations, the correlation CF is given by (for more details, see Refs.~\cite{Lisa:2005dd,Vidana:2023olz,Albaladejo:2024lam})
\begin{equation}
    C(\vec{p} \,) = \int \mathrm{d}^3 \vec{r}\, S_{12}(\vec{r} \,) |\psi(\vec{r}, \vec{p} \,)|^2,
\end{equation}
where $\psi(\vec{r}, \vec{p} \,)$ is the wave function of the two-particle system, and $\vec{p}$ and $\vec{r}$ represent the relative momentum and distance between two particles observed in the rest frame of the pair, respectively. $S_{12}(\vec{r} \,)$ is the source function, for which we consider a spherically symmetric gaussian profile:
\begin{equation}
    S_{12} (\vec{r} \, ) = \frac{1}{(\sqrt{4\pi})^3 R^3} \exp(-\frac{\vec{r}^{\,2}}{4R^2}),  
\end{equation}
where $R$ is a measure of the system size, typically in the range of $1$ to $5\,\text{fm}$. The wave function $\psi(\vec{r}, \vec{p})$ can be expressed in terms of the corresponding $T$-matrix elements and the modified loop function $\widetilde{G}$ (discussed below). Thus, the CF can be written as
\begin{eqnarray}
\mathcal{C}_{i} (p_i) &=& 1 + 4\pi\theta(\Lambda - p_i)\int_0^{\infty} \mathrm{d} r r^2\, S_{12}(r) \nonumber \\
&& \times\sum_j \omega_j^{(i)} \left\lvert j_0(p_i r)\delta_{ij} + T_{ij}(\sqrt{s}) \widetilde{G}_j(r,s) \right\rvert^2\,,
\end{eqnarray}
where $T_{ij}$ are the $T$-matrix elements for the coupled channel system, $p_i$ is the CM momentum of each coupled hadron pair, and $j_0(p_i r)$ is the zeroth-order spherical Bessel function. Additionally, the contributions from the $j$-th transitions are scaled by the so-called production weights $\omega_j^{(i)}$, which are computed following Appendix A of Ref.~\cite{Encarnacion:2024jge}, under the assumption of high-multiplicity events in p-p collisions. The production yields of the particles have been obtained employing the $\gamma_S$CSM model implemented in the Thermal-FIST package \cite{Vovchenko:2019pjl}, and the relative momentum distribution of each pair has been extracted from the single-particle momentum distributions given by the CBW model \cite{Schnedermann:1993ws}. The obtained values are shown in Table~\ref{Tab:weight}.
Above, the $\widetilde{G}_i(r, E)$ function is given by:
\begin{eqnarray}
    \widetilde{G}_i(r, s) &=& 2M_i \int_0^{\Lambda} \frac{q^2 \mathrm{d} q}{2\pi^2} \frac{\Omega_i(q)+E_i(q)}{2 \Omega_i(q) E_i(q)} \nonumber \\
    && \times \frac{j_0(q r)}{s-\left(\Omega_i(q)+E_i(q)\right)^2 + i\epsilon}\,,
\end{eqnarray}
where $\Omega_i(q) = \sqrt{m_i^2 + \vec{q}^{\,2}}$ and $E_i(q) = \sqrt{M_i^2 + \vec{q}^{\,2}}$. For numerical convenience, $\widetilde{G}_i(r, s)$ is evaluated following the prescription given in Eqs.~(7) and (8) of Ref.~\cite{Albaladejo:2023wmv}.

%%%%%%%%%
\section{Results}\label{Sec:res}

\begin{table}[!htbp]
\centering
\caption{Pole positions in physical basis for $S=-2, Q =-1$ sector.}
\setlength{\tabcolsep}{10pt}
\begin{tabular}{c|cc}
\hline\hline
$\Lambda\,(\text{MeV})$ & \multicolumn{2}{c}{$M_R - i \Gamma_R/2\,\text{(MeV)}$}  \\
& pole 1 & pole 2 \\ \hline\hline
\multicolumn{3}{c}{Without considering vector-meson widths} \\
\hline
$755$ & $1965.56 - i0$ & $2009.34 - i0.42$ \\
\hline
$795$ & $1950.05 - i0$ & $1994.12 - i0$ \\
\hline
$830$ & $1935.85 - i0$ & $1979.00 - i0$ \\
\hline
\multicolumn{3}{c}{Considering vector-meson widths} \\
\hline
$733$ & $1965.24 - i8.96$ & $2016.44 - i1.76$ \\
\hline
$775$ & $1949.79 - i7.06$ & $2000.91 - i0.84$ \\
\hline
$813$ & $1935.41 - i5.58$ & $1985.97 - i0.45$ \\
\hline\hline
\end{tabular}
\label{Tab:pole}
\end{table}

Given the current understanding of dynamically generated states from the HGS-based vector--baryon interaction \cite{Oset:2010tof,Gamermann:2011mq} and the available spectroscopic data in the the sector of $S=-2$, our strategy consists of matching the real part of the lowest generated pole, extracted from the scattering amplitudes, to the experimental mass of the $\Xi(1950)$ state ($M = 1950 \pm 15$~MeV) reported in the PDG compilation \cite{ParticleDataGroup:2024cfk}. This choice is motivated by two facts. On the one hand, the three-star status $\Xi(1820)$ is most likely dynamically generated in the pseudoscalar meson and decuplet baryon interactions \cite{Sarkar:2004jh,BESIII:2023mlv,Molina:2023uko}. On the other hand, the $\Xi(2030)$, which also has a three-star status, has $J \geqslant 5/2$ \cite{ParticleDataGroup:2024cfk} and thus cannot be generated with the $S$-wave dynamics of the present work. Therefore, the only remaining candidate with three-star status lying within the energy range explored in this manuscript is the $\Xi(1950)$. To accommodate this state in our amplitudes, we vary the cutoff value $\Lambda$ of Eq.~\eqref{eq:loop_hybrid} appropriately, considering two scenarios: with and without the inclusion of the finite widths of the $\olsi{K}{}^*$ and $\rho$ mesons. It is worth noting that, due to the assumptions made, all the states identified according to the previous criteria are degenerate in spin-parity, with ($J^P=1/2^-,3/2^-$). Moreover, from the group-theoretical point of view, the SU(3) decomposition into irreducible representations (irreps) for the coupling of the vector mesons of the $\rho$-nonet to the $1/2^+$ baryons of the nucleon octet reads~\cite{deSwart:1963pdg}: $(8\oplus1)\otimes8=1\oplus8\oplus8\oplus10\oplus10^*\oplus27\oplus8$. In the SU(3) basis $\ket{\phi;I;I_3;Y}$, where $\phi$ denotes the corresponding irrep and $Y=B+S$ is the hypercharge, the interaction is expected to be diagonal, except possibly within the subspace of the octets. Consequently, the eigenvalues of the matrix $C_{ij}$ indicate the attractive or repulsive character of each irrep. Then, inspecting the eigenvalues ($\lambda$) of the $I=1/2$ and $I=3/2$ coefficients once diagonalized\footnote{Note that, in the convention adopted in Ref.~\cite{Oset:2010tof}, as well as in the present work, a positive (negative) value of $C_{ii}$ indicates an attractive (repulsive) interaction.} (see Tables 11 and 12 in Ref.~\cite{Oset:2010tof}), one immediately identifies only two bound states from the two octets, with eigenvalue $\lambda=3$. These results support the existence of two $\Xi^*$ states in the $S=-2$ sector. 
 
In Table~\ref{Tab:pole} we present the poles for three different values of the cutoff $\Lambda$, chosen to fix the real part of the pole position at the central value and at the edges of the uncertainty range reported by the PDG compilation. Table~\ref{Tab:pole} displays the results for both cases: one that includes the finite width of the vector mesons, and another that neglects it. As expected from group theory, for each $\Lambda$ value, we find the $\Xi(1950)$ state (pole 1) and a second one, $\Xi^*$ (pole 2), around $2000$~MeV, both exhibiting the spin-parity degeneracy mentioned above. In particular, the central mass value of the $\Xi(1950)$ state~\cite{ParticleDataGroup:2024cfk} is pinned down for the cutoffs $\Lambda = 795 \mev$ and $\Lambda = 775 \mev$ in the cases without and with the inclusion of vector meson widths, respectively.\footnote{Note that the difference between these two cutoff values is only $20\,\text{MeV}$, much smaller than the considered vector-meson widths.} In Table~\ref{tab:couplings}, we present the couplings to different channels obtained for these two cases. The $\Xi(1950)$ state strongly couples to the $\rho \Xi$ channels, whereas the second $\Xi^*$ state predominantly couples to the \( \olsi{K}{}^* \Sigma \) channels. Note that by neglecting the widths of the \( \olsi{K}{}^* \) and \( \rho \) mesons, we obtain two bound states located approximately \( 60~\mathrm{MeV} \) and \( 15~\mathrm{MeV} \) below the \( K^{*-} \Lambda \) threshold. When the finite widths of the vector mesons are taken into account, the resulting states lie approximately \( 60~\mathrm{MeV} \) and \( 5~\mathrm{MeV} \) below the \( K^{*-} \Lambda \) threshold, with widths of about \( 15~\mathrm{MeV} \) and \( 2~\mathrm{MeV} \), respectively. This shift arises from the convolution of the \( \rho \) and \( \olsi{K}{}^* \) mass distributions in the loop function. By elimination, as discussed at the beginning of this section, the only reasonable assignment of a physical state to the second generated $\Xi^*$ pole is $\Xi(2120)$, as it is the only known state above $\Xi(1950)$ for which the available experimental information does not strongly preclude such an identification---recall that, as said earlier, there are strong indications that $\Xi(2030)$ has $J \geqslant 5/2$ \cite{Amsterdam-CERN-Nijmegen-Oxford:1977bvi}. This association was also proposed in Refs.~\cite{Oset:2010tof,Garzon:2012np}. In any case, the currently limited knowledge of $\Xi^*$ spectroscopy may be obscuring potential assignments to other, as yet unobserved, states that are expected to exist on symmetry grounds. Alternatively, the experimental signals or bumps may be concealing distinct, overlapping states, as in the case of the single experimental peak observed in the $K^-\Lambda$ invariant mass distribution from $\psi(3686) \to K^-\Lambda\bar{\Xi}^+$, which masks the double-pole structure of the $\Xi(1820)$. For simplicity, we will often refer to this second pole as $\Xi(2120)$ in what follows, although it is important to keep in mind the caveats discussed above.

\begin{table*}[!htbp]
\centering
\caption{Couplings to VB channels for the cases excluding (including) the vector widths, corresponding to $\Lambda = 795 \mev$ ($775 \mev$).
\label{tab:couplings}}
\setlength{\tabcolsep}{3.5pt}
{\small\begin{tabular}{cccccccc}
\hline\hline
 & $K^{*-} \Lambda$ & $K^{*-} \Sigma^0$ & $\rho^-\Xi^0$ & $\olsi{K}{}^{\ast0} \Sigma^-$  & $\rho^0 \Xi^-$ & $\omega \Xi^-$ & $\phi \Xi^-$ \\
\hline\hline
\multicolumn{8}{c}{Without considering vector-meson widths}\\ \hline
Pole 1 & $-1.80$ & $0.66$ & $\hphantom{+}2.76$ & $0.91$ & $\hphantom{+}1.95$ & $\hphantom{+}0.46$ & $-0.64$\\
\hline
Pole 2 & $\hphantom{+}0.62$ & $1.88$ & $-0.04$ & $2.68$ & $-0.05$ & $-1.47$ & $\hphantom{+}2.03$\\
\hline\hline
\multicolumn{8}{c}{Considering vector-meson widths} \\ \hline
Pole 1 & $-1.75 + i0.15$ & $0.62 - i0.09$ & $\hphantom{+}2.67 - i0.11$ & $0.85 - i0.06$ & $\hphantom{+}1.89 - i0.18$ & $\hphantom{+}0.46 - i0.03$ & $-0.64 + i0.04$ \\
\hline
Pole 2 & $\hphantom{+}0.55 - i0.11$ & $1.82 -i0.03$ & $-0.01 + i0.11$ & $2.60 - i0.03$ & $-0.03 + i0.07$ & $-1.43 + i0.05$ & $\hphantom{+}1.98 + i0.06$ \\
\hline\hline
\end{tabular}}
\end{table*}
\begin{figure*}[!htbp]
\begin{center}
\includegraphics[width=0.78\textwidth]{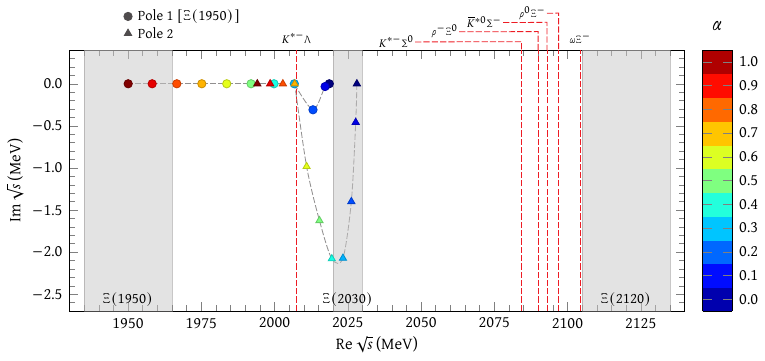}
\end{center}
\vspace{-0.7cm}
\caption{Evolution with the parameter $\alpha\!\in\!(0,1)$ of pole 1 [identified with $\Xi(1950)$ and coupled mostly to $\rho\Xi$] and pole 2 (coupling mostly to $\olsi{K}{}^{*} \Sigma$), represented with circles and triangles, respectively. The vertical dashed lines represent the thresholds of the channels appearing in the plotted range. The gray bands represent the masses, including uncertainties, of the three $\Xi$ states within the considered energy region.
\label{Fig:position}}
\end{figure*}
\begin{table*}[!htbp]
\centering
\caption{Values of the scattering length $a$ and effective range $r_0$ for the cases without and considering the widths of vector meson.
\label{tab:scatt_para}}
\setlength{\tabcolsep}{3.5pt}
{\small\begin{tabular}{cccccccc}
\hline\hline
 & $K^{*-} \Lambda$ & $K^{*-} \Sigma^0$ & $\rho^-\Xi^0$ & $\olsi{K}{}^{\ast0} \Sigma^-$  & $\rho^0 \Xi^-$ & $\omega \Xi^-$ & $\phi \Xi^-$ \\
\hline\hline
\multicolumn{8}{c}{Without considering vector-meson widths}\\ \hline
$a$ [fm] & $\hphantom{+}0.61 - i0.00$ & $\hphantom{+}0.32 - i0.00$ & $\hphantom{+}0.37 - i0.04$ & $\hphantom{+}0.42 - i0.03$ & $\hphantom{+}0.24 - i0.04$ & $\hphantom{+}0.18 - i0.03$ & $\hphantom{+}0.15 - i0.10$\\
\hline
$r_0$ [fm] & $-2.27 - i0.00$ & $-1.74 - i0.03$ & $-0.89 - i0.27$ & $-0.31 - i0.67$ & $-0.19 - i1.52$ & $-0.51 - i2.63$ & $\hphantom{+}0.27 - i0.37$\\
\hline\hline
\multicolumn{8}{c}{Considering vector-meson widths} \\ \hline
$a$ [fm] & $\hphantom{+}0.86 - i0.39$ & $\hphantom{+}0.38 - i0.09$ & $\hphantom{+}0.43 - i0.22$ & $\hphantom{+}0.53 - i0.15$ & $\hphantom{+}0.29 - i0.12$ & $\hphantom{+}0.20 - i0.04$ & $\hphantom{+}0.15 - i0.10$ \\
\hline
$r_0$ [fm] & $-8.45 - i0.70$ & $-4.28 -i0.64$ & $-3.21 - i0.36$ & $-3.76 - i0.30$ & $-3.44 + i0.62$ & $-1.55 + i1.38$ & $\hphantom{+}0.26 - i0.38$ \\
\hline\hline
\end{tabular}}
\end{table*}

To gain further insight into the nature of these states, it is interesting to study the dependence between pole the positions of the $\Xi(1950)$ and $\Xi(2120)$ states and the coupled channels considered. Here we take the case without considering the widths of the $\olsi{K}{}^\ast$ and $\rho$ mesons as an example. We keep the values of $V_{\olsi{K}{}^{\ast 0}\Sigma^- \to \olsi{K}{}^{\ast 0}\Sigma^-}$, $V_{\olsi{K}{}^{\ast 0}\Sigma^- \to K^{*-}\Sigma^0}$, $V_{\rho^-\Xi^0 \to \rho^-\Xi^0}$, and $V_{\rho^-\Xi^0 \to \rho^0\Xi^-}$, and multiply the other elements of Eq.~\eqref{eq:Vij} by a parameter $\alpha$, which ranges from $0$ to $1$. Therefore, the value $\alpha=0$ represents the case when only the channels from which the poles originate are kept, and $\alpha=1$ would correspond to our full amplitude. The evolution of the poles with $\alpha$ is shown in Fig. \ref{Fig:position}. It is remarkable to see that the mass of the state coupling mostly to $\rho^-\Xi^0$ $(\Xi(1950))$ is more sensitive to the parameter $\alpha$. It can be observed that both states have a similar behaviour. Pole 1 (pole 2) starts around $70\,\text{MeV}$ ($60\,\text{MeV}$) below the $\rho\Xi$ ($\olsi{K}{}^{\ast}\Sigma$) threshold. When $\alpha$ increases, \textit{i.e.}, when the other channels are progressively turned on, the masses of both poles decrease. They also acquire a finite width, as they can now decay into $\olsi{K}{}^\ast \Lambda$. When $\alpha$ is further increased, the masses drop below the $\olsi{K}{}^\ast \Lambda$ threshold, and then their widths vanish once again. Eventually, for $\alpha = 1$, the poles reach their final positions: 140~MeV and 90~MeV below the $\rho\Xi$ and $\bar{K}^{*}\Sigma$ thresholds for poles 1 and 2, respectively Note also that their coupling to the $\bar{K}^{\ast}\Lambda$ channel, which as discussed above is around $60\,\text{MeV}$ and $15\,\text{MeV}$ above poles 1 and 2, respectively, is sizeable, particularly for pole 1. All these coupled channel effects are relevant to understand the nature of these states, and partly explain their moderately large binding energies with respect to the main dynamical channels.

\begin{figure*}[t]
\begin{center}\begin{tabular}{lcr}
\includegraphics[width=0.33\textwidth]{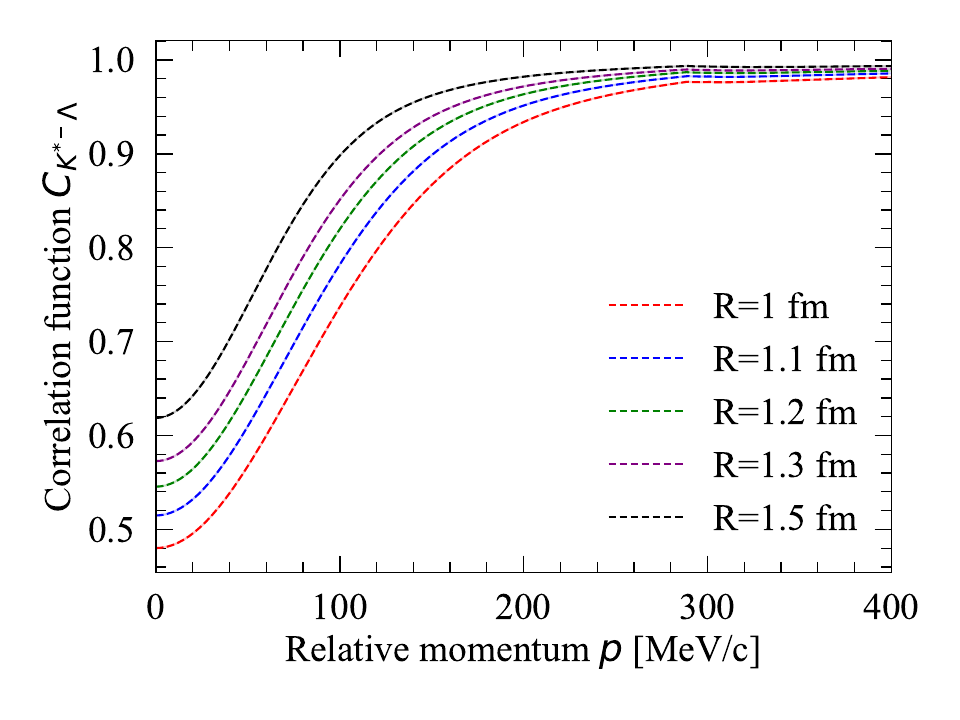} &
\includegraphics[width=0.33\textwidth]{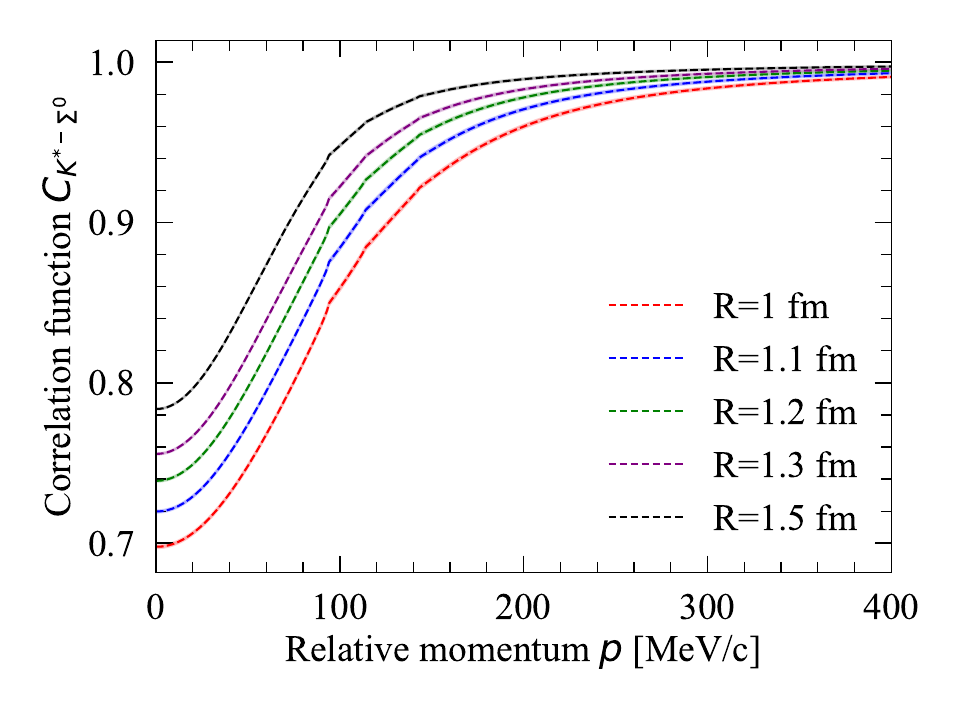} &
\includegraphics[width=0.33\textwidth]{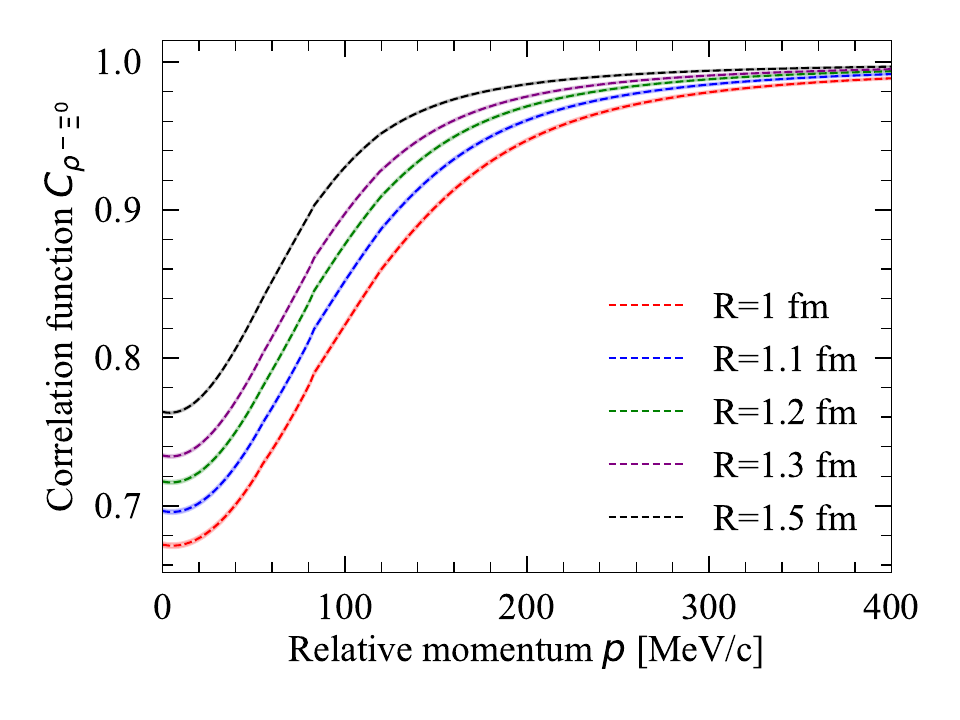}\\
\includegraphics[width=0.33\textwidth]{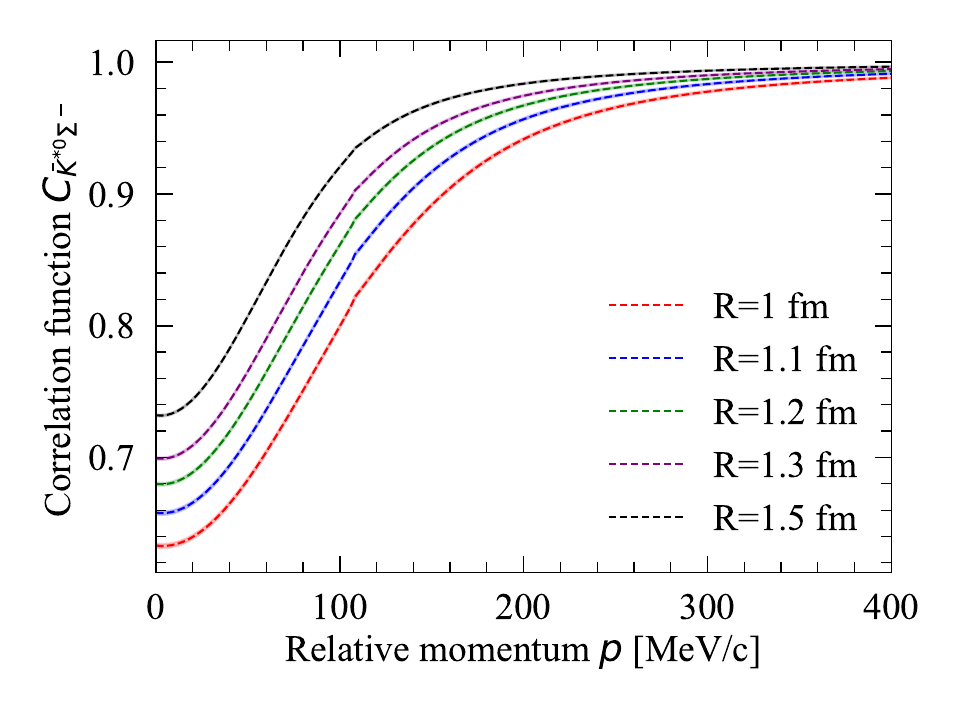} &
\includegraphics[width=0.33\textwidth]{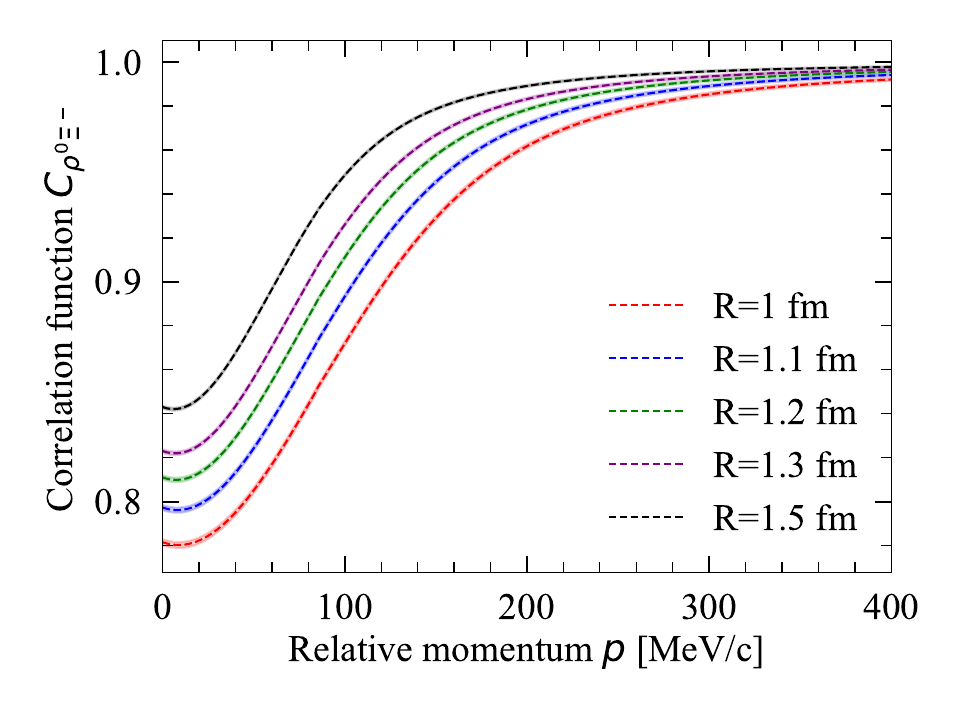} &
\includegraphics[width=0.33\textwidth]{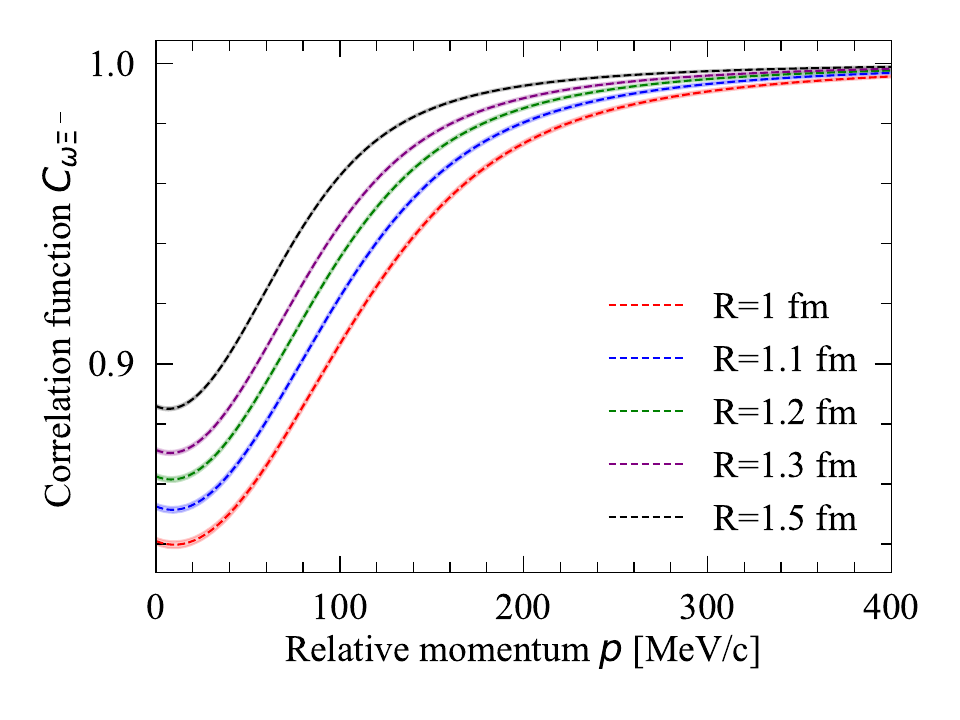}
\end{tabular}\end{center}
\vspace{-0.7cm}
\caption{Correlation function of the $K^{*-} \Lambda$, $K^{*-} \Sigma^0$, $\rho^-\Xi^0$, $\olsi{K}{}^{\ast 0}\Sigma^-$, $\rho^0\Xi^-$ and $\omega\Xi^-$ channels for different values of the source size, where the widths of $\olsi{K}{}^*$ and $\rho$ vector mesons are not considered. The 68\% CL color-shaded bands are estimated from the uncertainties in the production weights.}
\label{Fig:correlation_function_nowidth}
\end{figure*}

\begin{figure*}[t]
\begin{center}\begin{tabular}{lcr}
\includegraphics[width=0.33\textwidth]{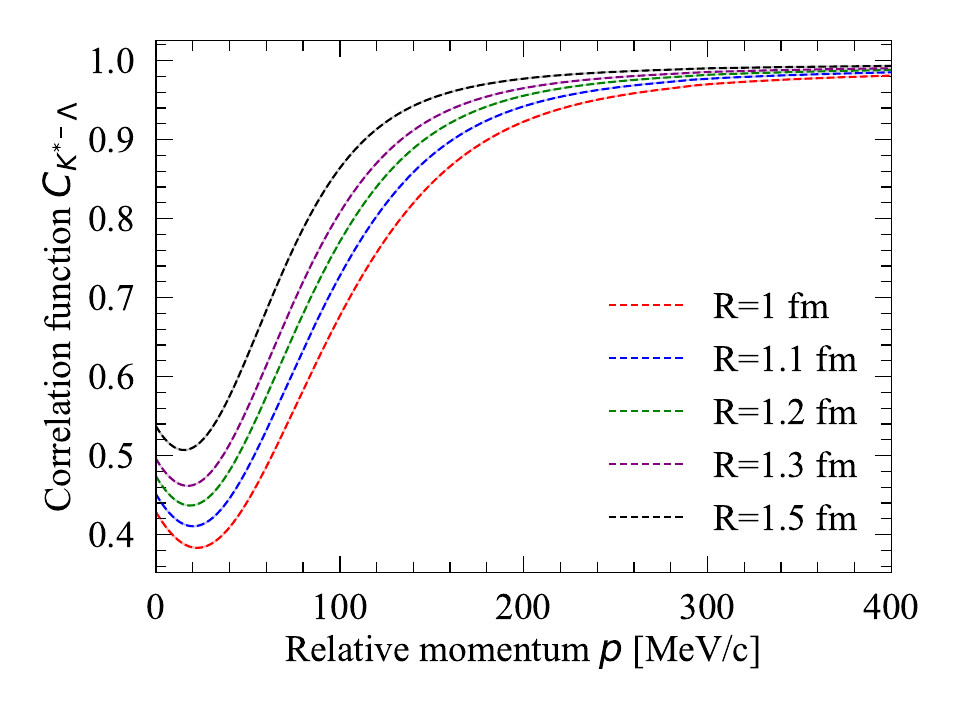} &
\includegraphics[width=0.33\textwidth]{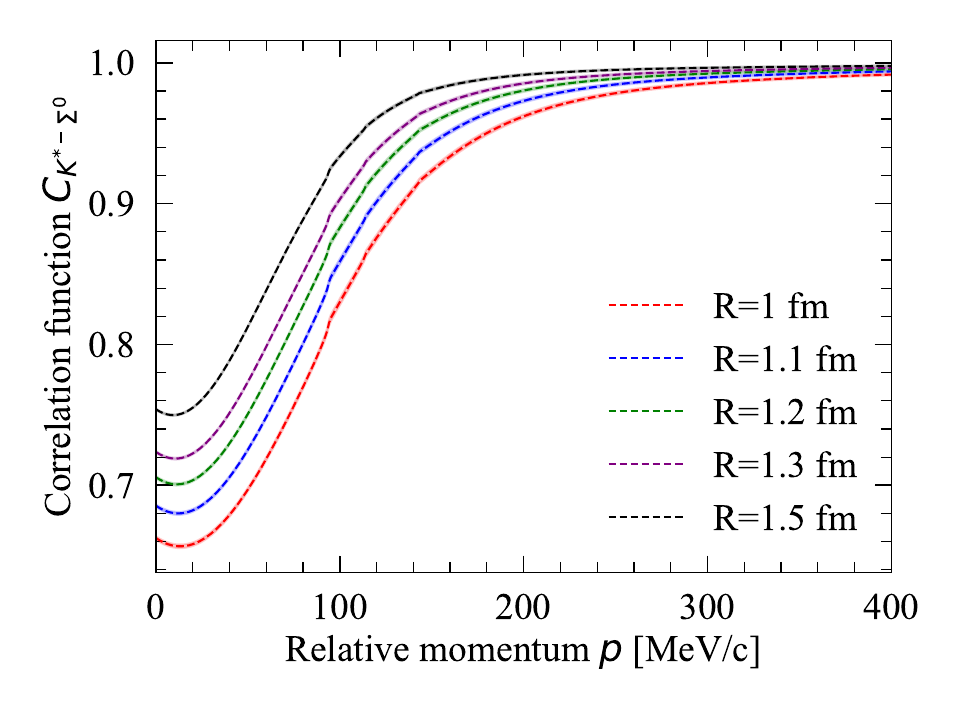} &
\includegraphics[width=0.33\textwidth]{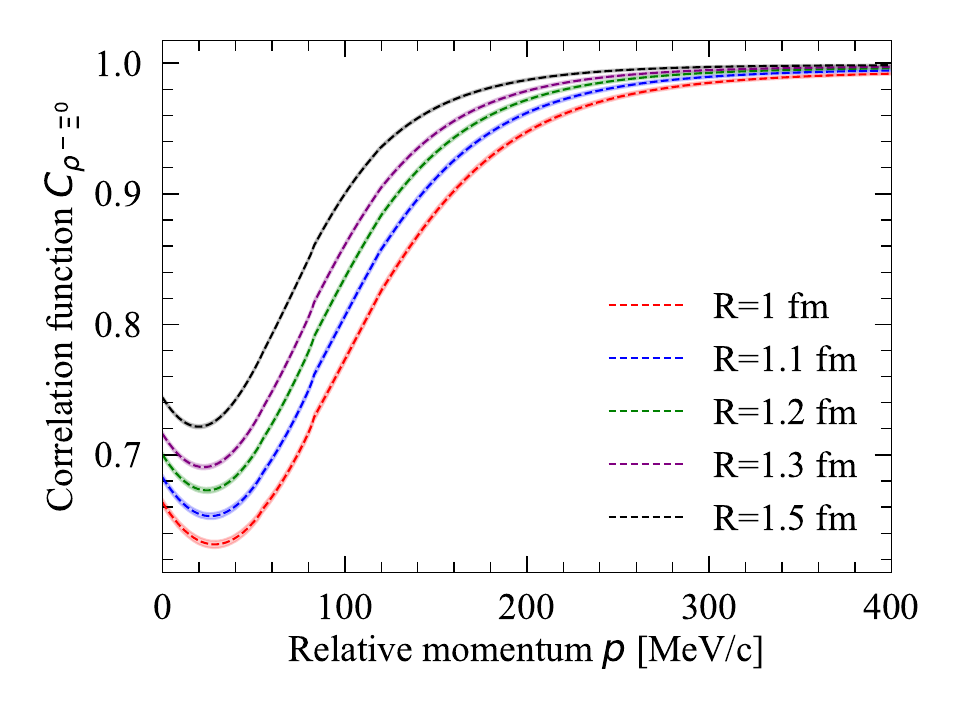} \\
\includegraphics[width=0.33\textwidth]{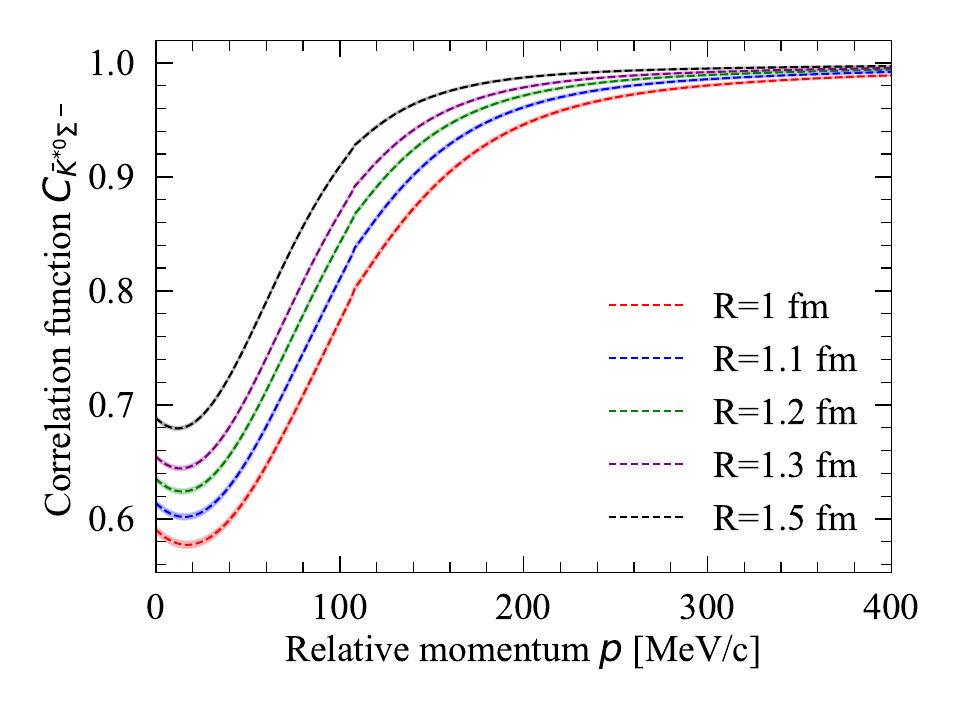} &
\includegraphics[width=0.33\textwidth]{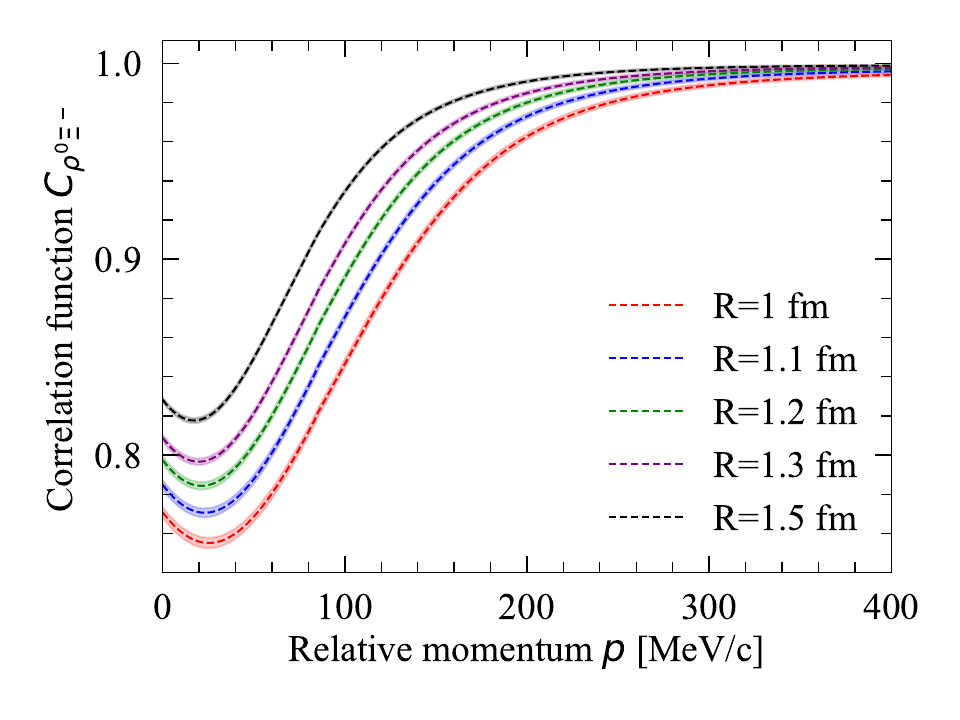} &
\includegraphics[width=0.33\textwidth]{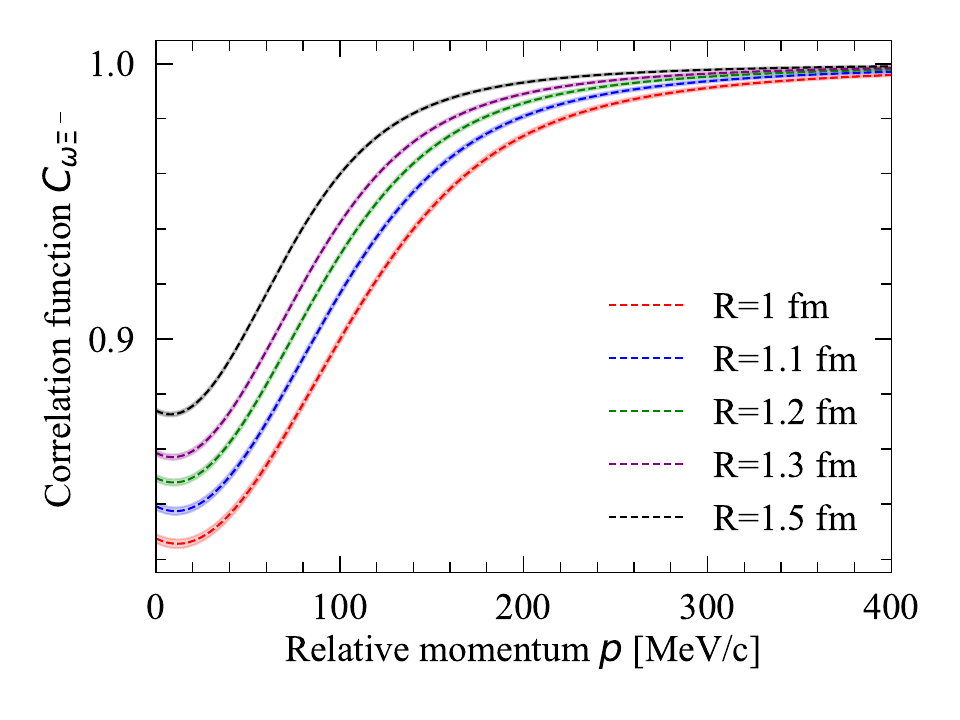}
\end{tabular}\end{center}
\vspace{-0.7cm}
\caption{Same as Fig.~\ref{Fig:correlation_function_nowidth}, but for the case where the widths of the $\olsi{K}{}^*$ and $\rho$ vector mesons are taken into account.}
\label{Fig:correlation_function_width}
\end{figure*}
In Table~\ref{tab:scatt_para}, we present the values of the scattering parameters, obtained using the expressions given in Eq.~(21) of Ref.~\cite{Encarnacion:2025luc}, where the effect of the vector-meson widths is already taken into account for the corresponding case. As an interesting result, it is worth mentioning that the dressing of the vector-mesons via convolution has a significant impact on the effective ranges \( r_0 \), due to the proximity of the first 6 thresholds, which involve both \( \olsi{K}{}^* \) and \( \rho \). Among the scattering lengths, three stand out as slightly larger: \( a_{\vphantom{\olsi{K}}K^{*-}\Lambda} \), \( a_{\olsi{K}{}^{*0}\Sigma^-} \), and \( a_{\rho^-\Xi^0} \). The enhancement of \( a_{\vphantom{\olsi{K}}K^{*-}\Lambda} \) can be attributed to the influence of nearby poles located just below the corresponding threshold. On the other hand, the larger values of \( a_{\olsi{K}{}^{*0}\Sigma^-} \) and \( a_{\rho^-\Xi^0} \) are mainly due to the direct contribution from the corresponding elastic transitions ($V_{ii} \neq 0$, see Table~\ref{tab:couplings}), unlike the other channels where the amplitude arises exclusively from coupled-channel effects. The obtained scattering parameters may serve not only as a useful benchmark for future experimental studies, but also as a valuable input for the discussion of the CFs presented below.

As discussed in the introduction, little is known about the baryon states in the $S=-2$ sector and, in this respect, experimental information coming from femtoscopy would be most welcome. For the reasons stated at the beginning of Sect.~\ref{sec:forma}, we restrict ourselves to the charge $Q=-1$ sector. In Figs. \ref{Fig:correlation_function_nowidth} and \ref{Fig:correlation_function_width}, we show our predictions, based on the chiral unitary approach followed in this work, for the CFs of the $K^{*-} \Lambda, K^{*-} \Sigma^0, \rho^-\Xi^0, \olsi{K}{}^{*0}\Sigma^-, \rho^0\Xi^-$ and $\omega\Xi^-$ channels for different values of the source size $R = 1, 1.1, 1.2, 1.3, 1.5\,\text{fm}$. The $\phi \Xi^-$ CF has been omitted due to the location of its threshold,  which lies far above the dynamically generated poles and, therefore, cannot provide any evidence of their existence. Fig.~\ref{Fig:correlation_function_nowidth} shows the CFs for the case in which the widths of the $\olsi{K}{}^*$ and $\rho$ mesons are not considered, whereas in Fig.~\ref{Fig:correlation_function_width} the vector-meson width effects are included. We can see that all CFs at the origin are smaller than 1. As the source size $R$ increases, each CF becomes larger and closer to 1 at a smaller momentum. According to Ref.~\cite{Liu:2023uly}, the overall shape of all former CFs can be attributed either to the presence of a near-threshold bound state or to the effect of a repulsive interaction near threshold. For the particular cases of $K^{\ast-}\Lambda$ and $\olsi{K}{}^{\ast0} \Sigma^{-}$, they experience an attractive effective interaction, driven by the nearby poles generated in the coupled-channel dynamics. In turn, this manifests in a positive scattering length for these channels (see Table\,\ref{tab:scatt_para}). In contrast, the positive scattering lengths of the other channels are related to the repulsive character of the effective interaction they experience.

To estimate the 68\% confidence-level (CL) band shown in all panels of Fig.\,\ref{Fig:correlation_function_nowidth}, a Monte Carlo sampling procedure is employed to propagate the uncertainties associated with the production weights listed in Table~\ref{Tab:weight}. The negligible size of these uncertainties relies on the fact that, in all the displayed CFs, the dominant contribution is that of the elastic channel, whose production weight has no uncertainty related to it. The largest variation of these CFs comes from the source radius $R$, as seen in Figs.~\ref{Fig:correlation_function_nowidth} and \ref{Fig:correlation_function_width}. Indeed, as observed in previous femtoscopic analyses~\cite{ALICE:2021njx,ALICE:2020ibs,ALICE:2023sjd}, the uncertainty in the source sizes $R_j$ constitutes the main contribution to the overall uncertainty of the CFs.

In Fig.\,\ref{Fig:correlation_function_width}, we show the CF of all possible channels for the case considering the widths of the vector mesons. We find that the results of these two scenarios are very similar. The main difference is the curvature of the CFs near threshold, which is more pronounced when the vector-meson widths are considered.
%

%%%%%%%%%
\section{Conclusions}\label{Sec:con}

We have studied the meson-baryon interaction in the $S=-2, Q=-1$ sector using the chiral unitary approach within the framework of the local hidden gauge formalism. In particular, we combined the cutoff and dimensional regularization methods in the loop function, keeping the good analytical properties of the latter. In this hybrid way, the loop function at threshold is determined by a cutoff value $\Lambda$. The $\Xi(1950)$ and $\Xi(2120)$ states with $J^P = \frac{1}{2}^-$ and $\frac{3}{2}^-$ are dynamically generated from the interactions of $K^{*-} \Lambda$, $K^{*-} \Sigma^0$, $\rho^-\Xi^0$, $\olsi{K}{}^{*0}\Sigma^-$, $\rho^0\Xi^-$, $\omega\Xi^-$, and $\phi \Xi^-$ channels. We discussed $\Xi(1950)$ and $\Xi(2120)$ states with two scenarios for the cases without and considering the widths of $\olsi{K}{}^*$ and $\rho$ mesons. The femtoscopic CFs of $K^{*-} \Lambda$, $K^{*-} \Sigma^0$, $\rho^-\Xi^0$, $\olsi{K}{}^{*0}\Sigma^-$, $\rho^0\Xi^-$, $\omega\Xi^-$, and $\phi \Xi^-$ channels have been calculated. With the results of the present paper, our aim is to encourage the measurement of these CFs in actual experiments. These experimental data would greatly help to better understand the nature of the $\Xi(1950)$ and $\Xi(2120)$ resonances and test the scenario proposed in this work. The dissimilarity in future comparisons between theoretical predictions and the corresponding experimental CFs may indicate either the need to incorporate additional elements into the employed dynamics or the presence of previously unseen states.

%%%%%%%%%

%\vskip150pt

\section*{Acknowledgments}

This work is supported by the Spanish Ministerio de Ciencia e Innovaci\'on (MICINN) under contracts PID2020-112777GB-I00, PID2023-147458NB-C21 and CEX2023-001292-S; by Generalitat Valenciana under contracts PROMETEO/2020/023 and  CIPROM/2023/59. M.\,A. acknowledges financial support through GenT program by Generalitat Valencia (GVA) Grant No.\,CIDEGENT/2020/002, Ramón y Cajal program by MICINN Grant No.\,RYC2022-038524-I, and Atracción de Talento program by CSIC PIE 20245AT019. J.X. Lin acknowledges the support of the China Scholarship Council. M. A. and A. F. thank the warm support of the ACVJLI.%

\bibliographystyle{apsrev4-1_MOD}
\bibliography{refs}
\end{document}